\documentclass[useAMS,usenatbib]{aa}              
\usepackage{amsmath,amssymb}
\usepackage{delarray}
\usepackage{graphicx}

\newcommand{\Msun}{\ensuremath{\mathrm{M}_{\odot}}}

\newcommand{\hmpc}{$\rm{h^{-1}Mpc}$ }

\newcommand{\cpr}{$c_{\rm pr}$}
\newcommand{\cch}{$c_{\rm ch}$}
\newcommand{\ctrue}{$c \,\,$}
\newcommand{\crsl}{$c_{\rm rsl}$}

\newcommand{\nicefrac}[2]{\leavevmode\kern.1em
            \raise.5ex\hbox{\the\scriptfont0 #1}\kern-.1em
      /\kern-.15em\lower.25ex\hbox{\the\scriptfont0 #2}}

\begin{document}

\title{Impact of the reduced speed of light approximation on the post-overlap neutral hydrogen fraction in numerical simulations of the epoch of reionization}


\author{P. Ocvirk \inst{1} \and D. Aubert \inst{1} \and J. Chardin \inst{1} \and N. Deparis \inst{1} \and J. Lewis \inst{1}}
\institute{Observatoire astronomique de Strasbourg, Universit\'e de Strasbourg, 11 rue de l'Universit\'e, 67000 Strasbourg, France\\
\email{pierre.ocvirk@astro.unistra.fr}
}
\date{Typeset \today; Received / Accepted}


\abstract
 {The reduced speed of light approximation is used in a variety of simulations of the epoch of reionization and galaxy formation. Its popularity stems from its ability to drastically reduce the computing cost of a simulation by allowing the use of larger and therefore fewer timesteps to reach a solution. This approximation is physically motivated by the fact that ionization fronts rarely propagate faster than some fraction of the speed of light. However, no global proof of the physical validity of this approach is available and possible artefacts resulting from this approximation therefore need to be identified and characterized to allow its proper use.}
{In this paper we investigate the impact of the reduced speed of light approximation on the predicted properties of the intergalactic medium.}
{To this end we used fully coupled radiation-hydrodynamics RAMSES-CUDATON simulations of the epoch of reionization.}
{We find that reducing the speed of light by a factor 5 (20, 100) leads to overestimating the post-reionization average volume-weighted neutral hydrogen fraction by a similar factor $\sim$5 (20, 100) with respect to full speed of light simulations. We show that the error is driven by the hydrogen - photon chemistry by considering the analytical solution for a strongly ionized hydrogen gas in photoionization equilibrium. In this regime, reducing the speed of light has the same effect as artificially reducing the photon density or the hydrogen photoionization cross section and leads to an underestimated ionizing intensity. We confirm this interpretation by running additional simulations using a reduced speed of light in the photon propagation module, but this time we keep the true speed of light in the chemistry module. With this set-up, the post-reionization neutral hydrogen fractions converge to the full speed of light value, which validates our explanation. Increasing spatial resolution beyond a cell size of 1 kpc physical, so as to better resolve Lyman-limit systems, does not significantly affect our conclusions.}
{}


\titlerunning{Reduced speed of light and the post-overlap volume-averaged neutral hydrogen fraction}

\maketitle

%




\section{Introduction}
The epoch of reionization (hereafter EoR) is the period during which the ultraviolet light of the first stars and galaxies ionized the intergalactic medium (hereafter IGM), in expanding HII regions of increasing size, finally merging together to produce the almost uniformly ionized IGM we see today. This process ended about 1 billion years after the Big Bang, and is tightly coupled to the formation and growth of galaxies during this period.
This period is the focus of a broad effort throughout the astrophysical community, which aims to understand in detail the timeline and identify the main sources contributing to the reionization of the Universe. In order to interpret the current data, and to prepare for the wave of new data expected from current and future redshifted 21 cm experiments (e.g. LOFAR, MWA, HERA, SKA) and upcoming space observatories such as JWST, accurate models of galaxy formation during the EoR are required. Such models also have implications for near-field cosmology, as it has been shown that the reionization history of the Local Group may affect the properties of its population of low-mass galaxies \citep{ocvirk2011,ocvirk2013,ocvirk2014,gillet2015}.

Modelling the EoR and its sources is a notoriously difficult problem; see \cite{dayal2018} for a review of a variety of early galaxy formation models and their large-scale effects. One the one hand, large boxes in the 100 Mpc or more are required in order to properly sample the variety of environments and rare large voids or early massive galaxies or galaxy clusters, such as in the cosmic dawn simulation suites \citep{codaI,codaII,codadom}. On the other hand, very high spatial resolution, down to the parsec scale, is necessary to describe the interstellar medium (hereafter ISM) of galaxies, and the complex processes the ISM drives and reacts to such as star formation, supernova feedback, and the ionizing escape fraction \citep{geen2016,trebitsch2017,gavagnin2017,butler2017,costa2018}. The combination of large volume and high spatial resolution makes numerical simulations of galaxy formation during the EoR extremely challenging in principle and computationally costly with the rich set of physics required.

In particular, ionizing radiation is a key process in these simulations, both for its impact in reionizing the Universe and for the additional feedback it provides, and is especially costly to model. Two main classes of solutions have been adopted for treating hydrogen-ionizing radiation. Ray-casting algorithms involve sampling the radiation field produced by each source by a Monte Carlo technique and following each photon packet through the computational domain and its interaction with the hydrogen gas \citep{licorice,baek2009,traphic}. Another popular solution consists in considering the photons as a fluid and using the M1 closure relation \citep{levermore1984,dubroca1999,ripoll2001,audit2002,gonzalez2007,aubert2008,aubert2010}. This approach has led to the development of fully coupled radiation-hydrodynamics (hereafter RHD) galaxy formation codes, such as RAMSES-CUDATON \citep{codaI,codaII}, EMMA \citep{aubert2015}, and RAMSES-RT \citep{rosdahl2013}. In these codes, the radiative timestep is set by the Courant condition and is typically 100-1000 times shorter than the hydrodynamical timestep of the simulation. This simply reflects the fact that light propagation is much faster than any other process in this framework. This alone makes full RHD simulations very expensive. While some authors have alleviated this problem by accelerating the radiative transfer (hereafter RT) module using GPUs\footnote{Graphics Processing Units} \citep{aubert2010,aubert2015,codaI,codaII}, others have resorted to the so-called reduced speed of light approximation (hereafter RSL), in order to bring the radiative timestep closer to the hydrodynamic timestep and reduce the overhead due to RT \citep{kimm2013,rosdahl2013,codadom}.
This can drastically reduce the RT computing time. As a consequence, RSL is extremely popular and has allowed groups to perform simulations which would be impossible using the true speed of light. Using RSL may appear justified in dense regions (in particular in the ISM), where the ionization front (I-front) is much slower than the real speed of light. However, beyond this, no clear proof of the global physical validity of this approach has ever been provided. For instance, in optically thinner regions of average cosmic density or voids, ionization fronts may be faster than the reduced speed of light used, and the timing and geometry of reionization could be affected \citep{bauer2015,gnedin2016,deparis2019}. More recently, \cite{katz2017,katz2018} and \cite{SPHINX2018} have introduced and deployed the variable speed of light formalism with adaptive mesh refinement simulations, in which the speed of light is divided by a factor of 2 at each additional spatial refinement level. This allows us to keep the speed of light close or equal to its true value on the base grid, where ionization fronts move fast, while making it small enough to be computationally manageable in the dense, highly refined regions. This approach may mitigate some of the shortcomings of RSL, in particular with respect to the front speeds in low density regions. 
However, the validity of RSL nor variable speed of light with respect to predicting the {neutral hydrogen fraction}, has not been studied much before in RHD cosmological simulations, and this is what we aim to investigate in the present paper. In Sec. \ref{s:results}, we present simulations of the EoR performed with RAMSES-CUDATON using 100\%, 20\%, 5\%, and 1\% of the speed of light, and show that RSL simulations systematically overpredict the post-reionization neutral hydrogen fraction. We then interpret this result using analytical arguments and further simulations. We finally close the paper with our conclusions.

\section{Methodology}
\label{s:methodology}
\subsection{Simulations}
We used RAMSES-CUDATON \citep{codaI}, a coupling between RAMSES \citep{teyssier02} and ATON/CUDATON \citep{aubert2008,aubert2010}. Thanks to the CUDA optimization of ATON, RAMSES-CUDATON is able to take advantage of GPU acceleration and routinely performs RHD simulations with the full speed of light. It is therefore an adequate testbed for investigating artefacts in RSL simulations, which are less computationally expensive. The set-up used in this work is a 4 \hmpc box discretized on a fixed $256^3$ grid with parameters similar to the Cosmic Dawn I simulation performed on Titan at Oak Ridge National Laboratory \citep{codaI}, but with several differences which we detail here: (i) The cosmology adopted is compatible with Planck 2016 \citep{planck2016}. 
We adopted $\Omega_\Lambda=0.693$, $\Omega_m=0.307$, $\Omega_b=0.048$, $h=0.677$, and the power spectrum is described by $\sigma_8=0.8288$, with an index $n=0.963$; (ii) the fiducial star formation efficiency is $\epsilon_{\star}=0.033$, promoting earlier reionization; (iii) the stellar particle ionizing radiation escape fraction is $f_{\rm esc}=0.1$; and (iv) the temperature criterion for star formation has been removed, allowing star formation in all cells with a gas density higher than 50 times the average baryon density $\delta_{\star}$=$50 \, \langle \rho_b \rangle$. In CoDa II \citep{codaII}, similar parameters allowed us to reproduce  a number of observable constraints of the EoR. Therefore we consider these parameters to be adequate for the purpose of this study. For our 4 \hmpc box, the resulting reionization history is shown as the thick black line in Fig. \ref{f:RSL}. Although the box successfully reionizes around z=6.5 for our full speed of light run, the timing of reionization is not of great importance for this experiment, since we are more interested in investigating the post-reionization ionized fraction when varying the speed of light.
  \subsection{Reduced speed of light, dual speed of light}
  The ATON RT module consists of two main sub-modules: an advection (i.e. photon propagation) module and a thermochemistry module, which is purely local and models the interaction between the photons and the hydrogen gas. We therefore consider the two following speeds of light:
 {\cpr: the speed of light used in the propagation module; and}
{\cch: the speed of light used in the thermochemistry module.}
  With these definitions, the usual RSL approximation is obtained by using the same reduced speed of light \crsl=\cpr=\cch in both modules. This is the approach of \cite{rosdahl2013}, \cite{trebitsch2017}, and \cite{katz2017} for instance. As we will see, this has a strong impact on the volume-weighted average neutral ionized fraction after reionization.
{We note that this methodology differs from that of \cite{gnedin2016}, who distinguishes between the radiation background and the fluctuations on top of it. In the latter case the reduced speed of light is only used to propagate the fluctuations. We caution that the results presented in this work may not apply to this formulation.}
  
We also introduce the dual speed of light approximation (hereafter DSL), where the speed of light \cpr is reduced in the propagation module,  but maintained to its true value  \cch=\ctrue in the chemistry module. We performed one reference full speed of light simulation with \cpr=\cch=\ctrue, three RSL simulations with \cpr=\cch$=0.2 c$, $0.05 c$, and $0.01 c$, respectively. And finally we performed three DSL simulations with the true speed of light \cch=1 in the chemistry module but we varied the propagation speed \cpr=$0.2 c$, $0.05 c$ and $0.01 c$. The latter helps us understand the shortcomings of the RSL runs. Finally, a resolution study was also performed. The results are presented in the next section.

\section{Results}
\label{s:results}

\begin{figure}[t]
  \includegraphics[width=1.15\linewidth,clip]{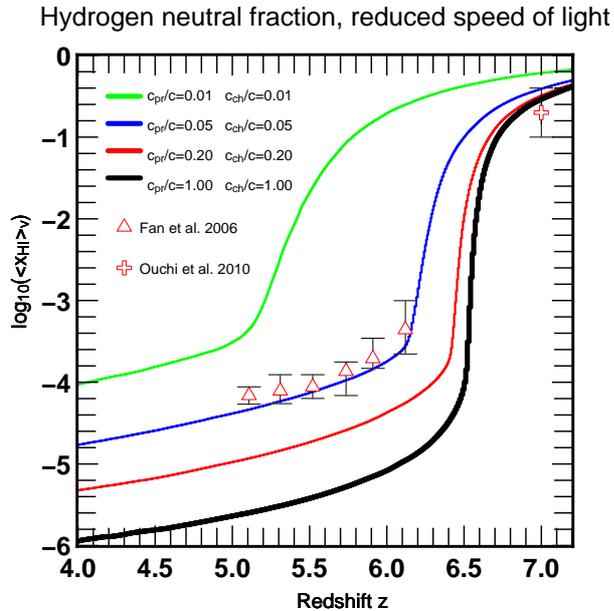}

  \caption{Reionization histories (volume-averaged neutral hydrogen fraction) of our RAMSES-CUDATON simulations in the RSL approximation. }
    \label{f:RSL}
\end{figure}

\begin{figure}
  \includegraphics[width=0.9\linewidth,clip]{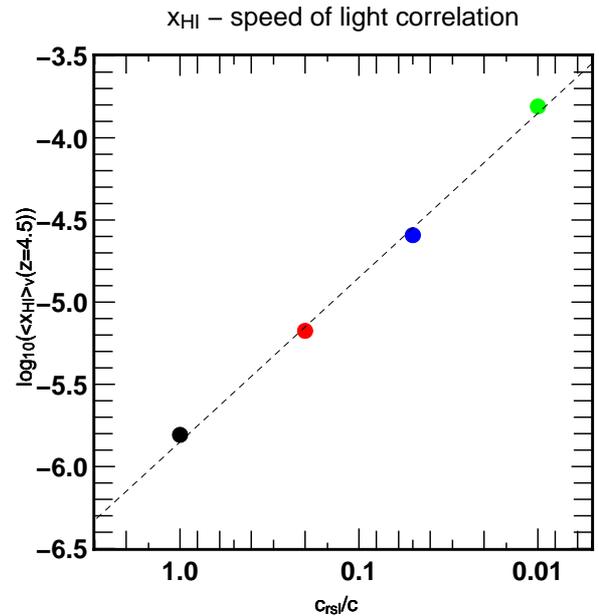}
  \caption{Correlation between the post-overlap volume-weighted neutral fraction (taken at z=4.5) and the adopted speed of light in our RSL simulations. The dashed line represents a linear correlation of slope 1.}
  \label{f:xHIvsc}
\end{figure}

\begin{figure*}[h]
\includegraphics[width=0.5\linewidth]{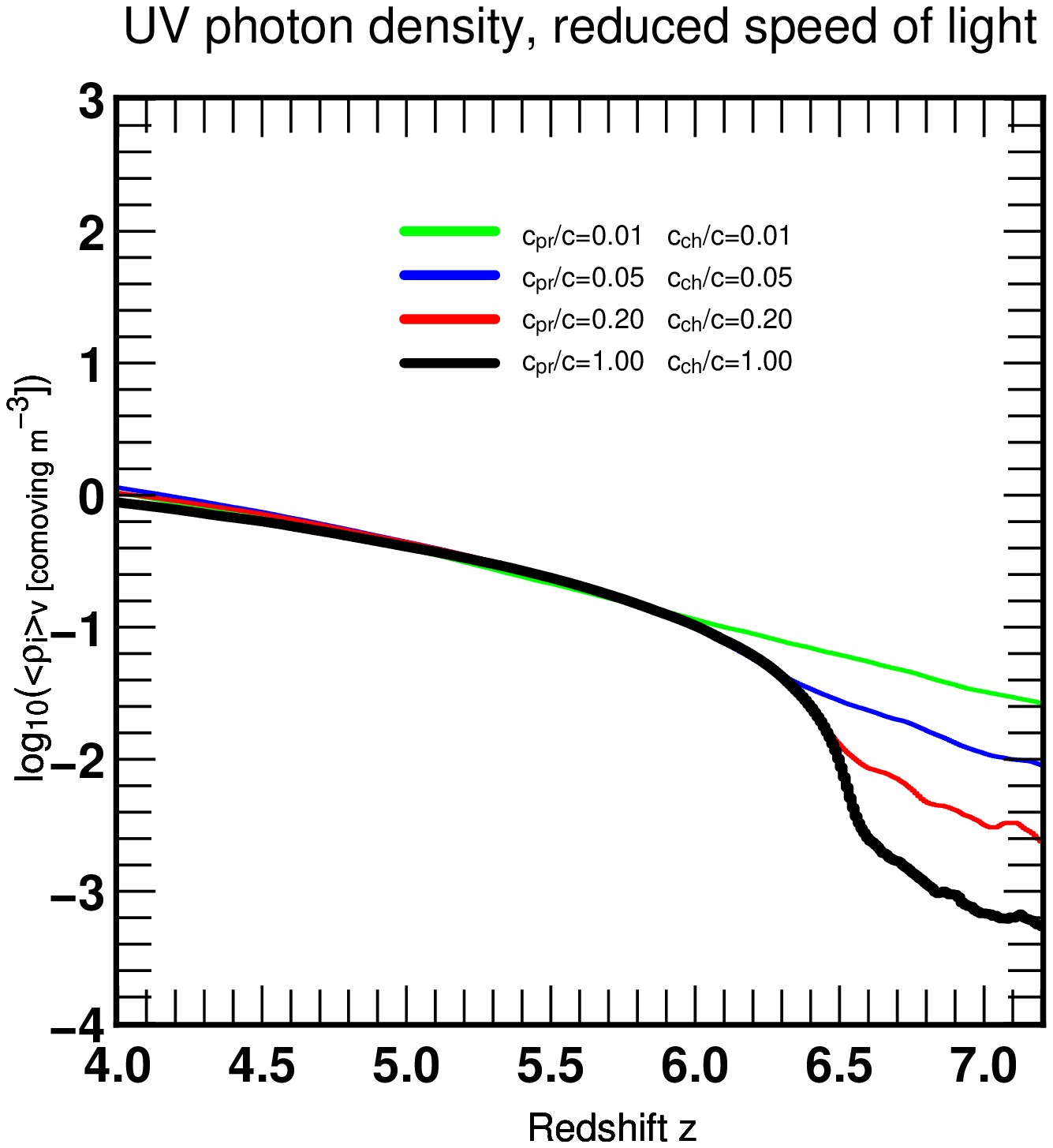}
\includegraphics[width=0.455\linewidth]{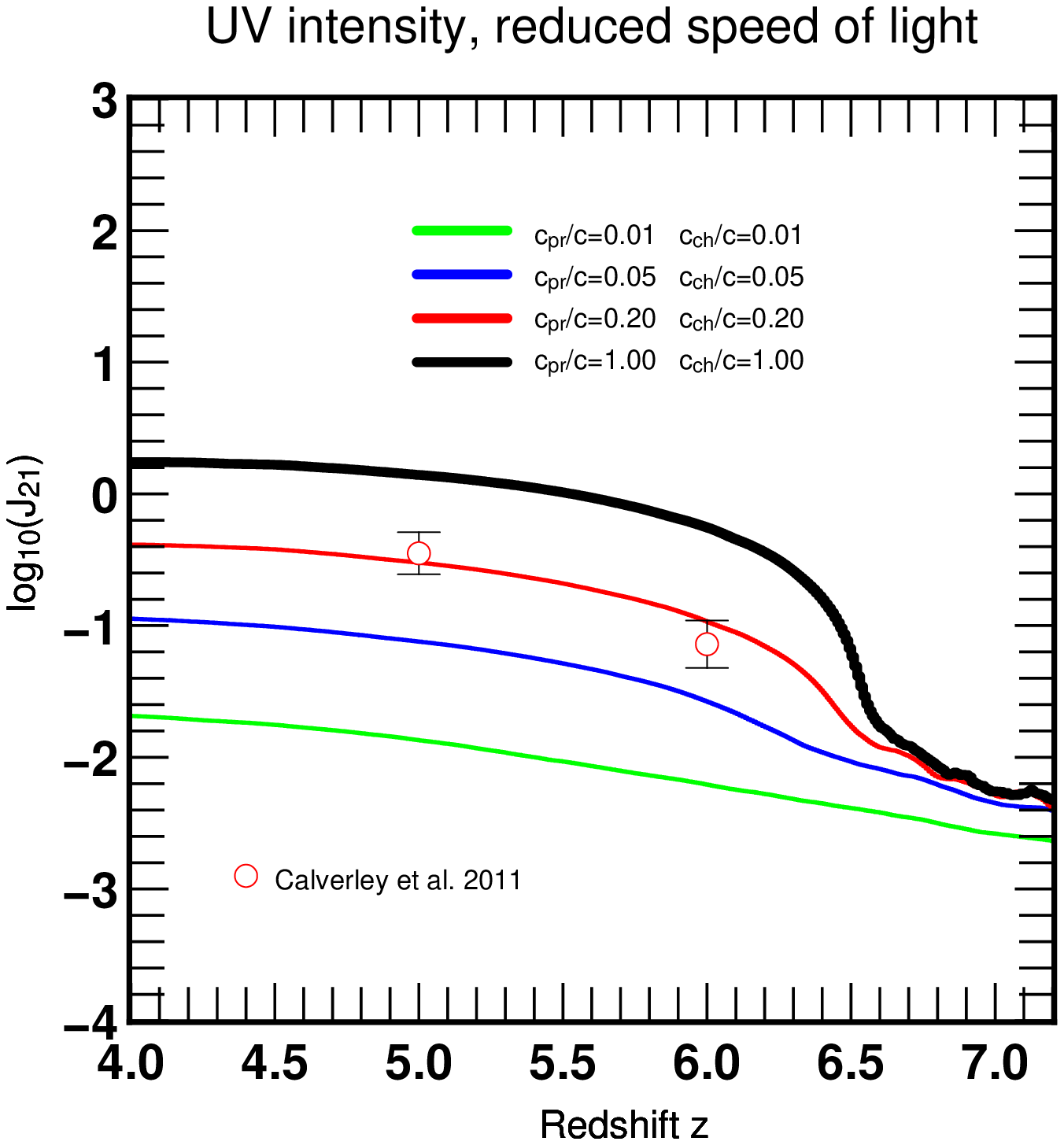}
\caption{Volume-averaged ionizing photon density ({ left}) and ionizing intensity ({ right}) for our 4 RSL runs. We note how similar post-reionization ionizing photon { densities} between the runs translate into different ionizing { intensities} due to the RSL approximation.}
\label{f:photd}
\end{figure*}

In this section we examine the impact of the RSL approximation on the post-reionization volume-averaged neutral fraction for different values of the speed of light.
\subsection{Reduced speed of light}
The reionization histories obtained in the RSL approximation with \crsl=\cpr=\cch=$c$, $0.2 c$, $0.05 c,$  and $0.01 c$  are shown in Fig. \ref{f:RSL}. The end of reionization, marked by a sudden drop in neutral fraction, is clearly delayed with respect to the full speed of light solution (thick black line), although for \crsl=$0.1 c$   the lag is rather small. These temporal aspects will be studied in more detail in \cite{deparis2019}. In this work, instead, we focus on the neutral hydrogen fraction. First of all, our run with the full speed of light yields a neutral fraction in disagreement with the observations of \cite{fan2006}. This is actually commonplace among simulations using the full speed of light \citep{aubert2010,petkova2011,so2014,bauer2015,codaI,codaII}. We find that the best agreement with \cite{fan2006} is found for the $0.05 c$ run. This shows that obtaining the correct neutral fraction after overlap in RSL does not guarantee that the correct answer will still be obtained when using the full speed of light. However, this does not necessarily rule out the use of RSL if simulations with different speeds of light converge on a neutral fraction for a given reionization history.

More importantly, we find that the post-reionization average neutral fractions are spread in a wide sequence parametered by the speed of light, where low speeds of light yield later reionization and higher neutral fraction.




In all RSL runs, the neutral fraction after reionization is too high compared to the full speed of light solution, by a factor $\sim 5$ ($\sim 20$, $\sim 100$) for \crsl=$0.2 c$ (\crsl=$0.05 c$, \crsl=$0.01 c$  respectively). Indeed, this strong quantitative correlation between the adpoted speed of light and the post-overlap neutral fraction is shown in Fig. \ref{f:xHIvsc}. The post-overlap fractions are taken at z=4.5, where all simulations have completed reionization. The points corresponding to our four fiducial RSL runs form a sequence that is well represented by a linear correlation of slope 1 (dashed line), suggesting that \crsl is the main parameter governing the post-overlap neutral fraction in these simulations. We can understand this by considering the analytical expression for the neutral fraction of a hydrogen gas in photoionized equilibrium, in the strongly photoionized limit (i.e. $x_{\rm HI}<0.01$ for instance). In this case, the neutral fraction behaves as\begin{equation}
  x_{\rm HI} \sim \alpha_B \rho_{\rm H}/c \rho_i \sigma \,
  \label{eq:xHI}
,\end{equation}
where $\alpha_B$ is the case B recombination coefficient, $\rho_{\rm H}$ is the total hydrogen density, $\rho_i$ is the ionizing photon density, $\sigma$ is the ionizing cross section of hydrogen, and $c$ is the speed of light. This expression shows that, in this regime, the neutral hydrogen fraction follows an inverse linear dependence with the speed of light $c$: dividing $c$ by a factor 5 results in a neutral fraction increased by the same factor 5 with respect to the full speed of light solution, and this is indeed what we find in the RSL simulations shown in Fig. \ref{f:RSL}.
The other terms in this expression are subject to only small variations between our simulations. Indeed, the case B recombination coefficient $\alpha_{\rm B}(T)$ is a function of temperature, which in our post-reionization IGM is the temperature of post-reionization photoionized hydrogen, i.e. 5 000 - 20 000 K. Moreover, the average post-reionization ($z=5$) ionizing photon density is about $\rho_i \sim 0.5$ photon per comoving m$^3$ for all RSL runs, as shown in the left panel of Fig. \ref{f:photd}. 
These two parameters are therefore not likely to cause the order of magnitude offsets seen in the RSL runs.

This trend of RSL simulations yielding higher neutral fractions with respect to full speed of light simulations is also seen in the literature, although it has not been clearly identified, explained, and reported until now. For instance, this result is similar to the $z<6$ region of Fig. 13 of \cite{bauer2015}, hereafter B15, although the authors only comment on the lagging reionization history of their RSL runs.

\begin{figure}[t]
  \includegraphics[width=0.9\linewidth]{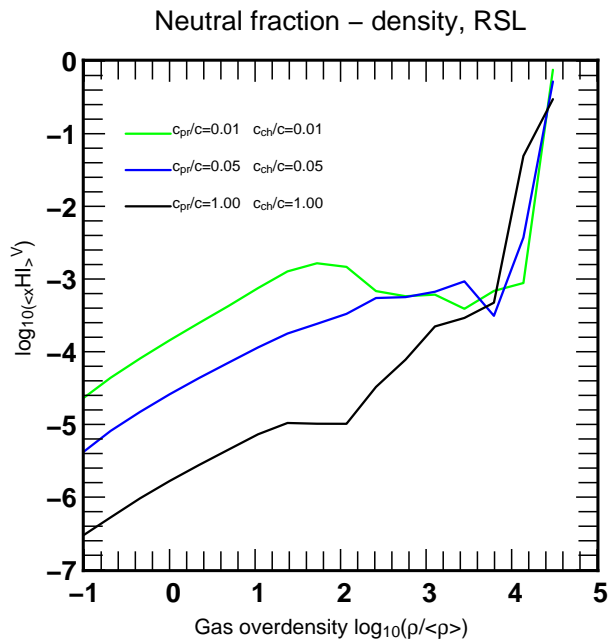}
  \caption{Average neutral fraction as a function of gas density for RSL and full speed of light runs at z=4.4.}
  \label{f:cobraRSL}
\end{figure}

Not all regions are equally impacted by the change of speed of light. This can be verified using Fig. \ref{f:cobraRSL}, where we show the ionized fraction of our simulations as a function of gas overdensity for the z=4.4 snapshot. The neutral fractions converge for overdensities larger than $\sim 1000$. 


\subsection{Attempt at reconciling RSL simulations with the full speed of light run}

\begin{figure}[t]
  \includegraphics[width=1.1\linewidth]{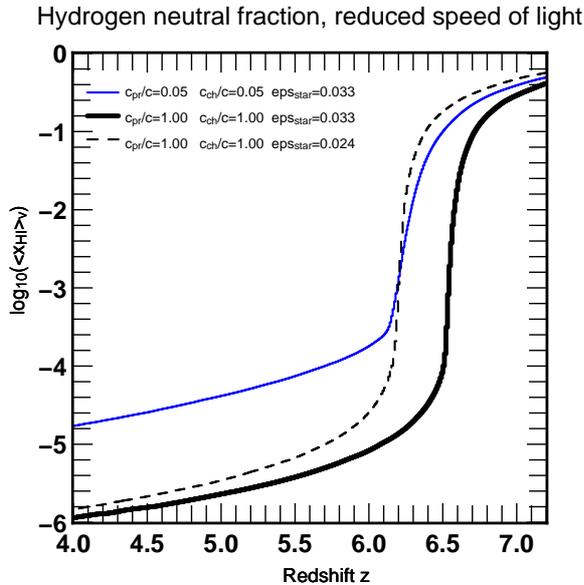}
  \caption{Reionization histories of our fiducial \crsl/$c$=0.05 and \crsl/$c$=1 runs and a \crsl/$c$=1 run with lower star formation efficiency. The lower star formation efficiency delays reionization, but still yields a post-overlap neutral fraction far below that of the RSL run.}
  \label{f:RSLretuned}
\end{figure}

Could it be that the higher neutral fraction of the RSL runs is a consequence of later reionization? Structures would be less ionized because they have spent a shorter time seeing the background. In other terms, is it possible to retune a full speed of light simulation (\cpr=\cch=$c$) to yield the same outcome as the fiducial \crsl=$0.05 c$ RSL run?
To investigate this aspect we ran a full speed of light simulation with a lower star formation efficiency $\epsilon_{\star}=0.024$ than our fiducial models to make reionization happen close to our fiducial \crsl=0.05 run. The result is shown in Fig. \ref{f:RSLretuned} (dashed line). The reionization history we obtained essentially looks like a delayed version of our fiducial $c=1$ run, horizontally offset by a $\Delta z=0.7$ to later times. Although this run reionizes at a similar time compared to the fiducial \crsl=0.05 run, it still yields a much lower post-overlap neutral fraction. This shows that the offsets in post-overlap neutral fractions between our fiducial runs are not due to offsets in the timing of reionization: even if we retune our runs (via e.g. the star formation efficiency as we have done in this example) so as to achieve the same reionization redshift for all \crsl, this can not compensate the offsets in post-overlap neutral fractions.
In Sec. \ref{s:res}, we explore another parameter with a potential impact on the neutral fraction, i.e. spatial resolution.

\subsection{Dual speed of light}
\begin{figure}[t]
  \includegraphics[width=1.1\linewidth]{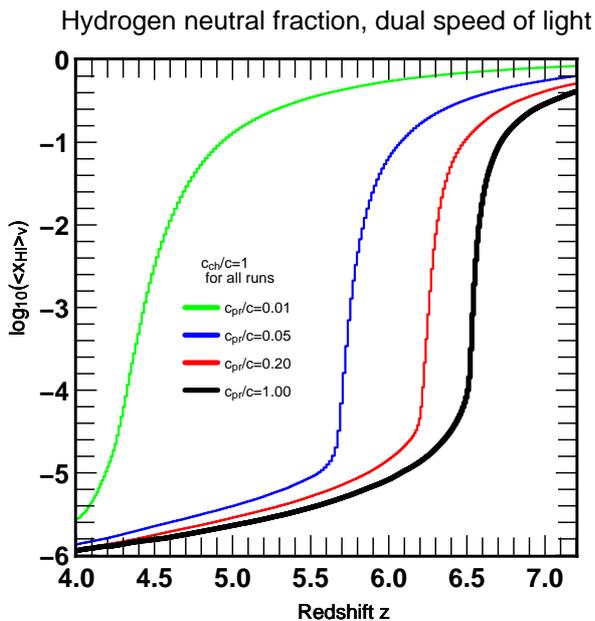}
  \caption{Reionization histories of our dual speed of light runs.}
  \label{f:DSL}
\end{figure}


It appears from Eq. \ref{eq:xHI} that the large error in post-reionization equilibrium fraction obtained with RSL may be entirely due to the use of a reduced speed of light in the thermochemistry solver, which is similar to articifically reducing the reaction cross section or the photon density. We confirm this explanation with three additional simulations, using a reduced speed of light in the propagation module \cpr=$0.2 c$, $0.05 c$,  and $0.01 c$, but keeping this time the true speed of light \cch=$c$ in the chemistry module. This set-up assumes simultaneously two different speeds of light, hence the term DSL approximation.


The resulting reionization histories are shown in  Fig. \ref{f:DSL}. In stark contrast to the RSL runs, the average neutral fractions after reionization converge this time to the full speed of light solution, clearly demonstrating that the reduced speed of light in the chemistry solver is indeed the origin of the large overestimate found with RSL.

Although the DSL approximation produces the correct post-reionization neutral fraction, it comes with its own set of caveats and artefacts. For instance, a slower propagation speed artificially increases the photon density around sources compared to the true speed of light case. In RSL, this is exactly compensated by a reduced speed of light in the photoionization rate, and this is why RSL yields correct neutral fractions for Stromgren spheres, as shown by \cite{rosdahl2013}. Conversely, in DSL, where the photoionization rate uses the true speed of light, the increased photon density produces over-ionized Stromgren spheres. In our runs, this in turn leads to an enhanced star formation rate (SFR) suppression in galaxies. Both aspects are illustrated in Appendix \ref{s:overi} and \ref{s:sfrrsldsl}.

For these reasons, we cannot at this stage advocate for the use of DSL as a replacement of RSL.
It seems RSL works best before overlap, while DSL is preferrable after overlap. Ideally, we would need to switch from one framework to the other along the course of a simulation. A physically motivated scheme for switching from RSL to DSL as a function of time and the radiative state of a region would be highly desirable but is beyond the scope of this paper.

\subsection{Resolution study}

\label{s:res}

\begin{figure}[t]
  \includegraphics[width=1.\linewidth]{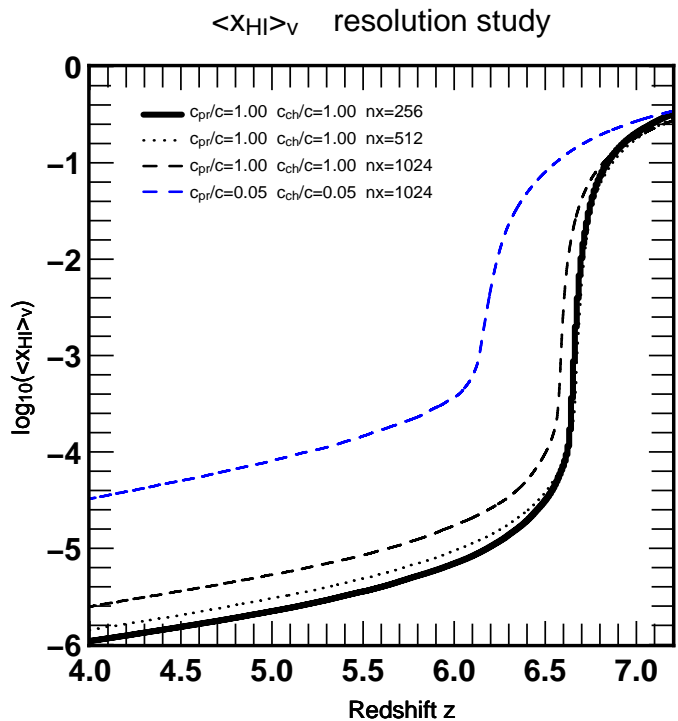}
    \caption{Reionization histories of 4 \hmpc boxes simulations with spatial resolutions of 1,2, and 4 times our fiducial resolution. The fiducial run is shown as a thick black line. All runs were calibrated so as to reach x$_{HI}=0.5$ between z=6 and z=7.
      The detailed properties of these simulations are listed in Tab. 1.} 
    \label{f:res}
\end{figure}


It is commonly accepted that resolving Lyman-limit systems (hereafter LLSs) is necessary to describe properly the population of absorbers in the IGM and circum-galactic medium  and predict the correct average neutral hydrogen fraction \citep{miralda-escude2000}. Therefore spatial resolution is expected to have some impact on the post-overlap neutral fraction. 
In particular, \cite{rahmati2018} quoted a physical size of $1-10$ kpc for LLSs, and claimed that therefore a spatial resolution of sub-kiloparsec physical is required to describe these systems. While our fiducial simulations satisfy this constraint at z$\sim 20$, they do not satisfy it after overlap: at z=5.5, their cell size is about 3.85 physical kpc.
Therefore we ran an additional suite of three simulations to check the impact of physical resolution on our results, where we varied the spatial resolution in a factor of 2 and 4, respectively, with respect to our fiducial simulation. This also yields a mass resolution increase of a factor 8 and 64, respectively.
The reionization histories of these simulations are shown in Fig. \ref{f:res}. These new simulations achieve complete reionization between redshifts 6 and 7. However, because the effect we are after ends up being rather small, for a better comparison we shifted their reionization histories to match the reionization redshift of the fiducial run. With this shift applied, they line up perfectly during the overlap period, but produce different tracks after overlap. The neutral fractions of higher spatial resolution simulations are higher than in our fiducial run: each factor of 2 increase in spatial resolution results in an increase in neutral fraction after overlap of about $\sim \times 1.4 - 1.6$ with respect to the fiducial run. The exact numbers are reported in Tab. 1.

\begin{table*}[t]
  \label{t:resstudy}  
  \begin{center}
    \begin{tabular}{lcccccccc}
      \hline
    Simulation & \cpr $/c$ & \cch $/c$ & N$_{\rm DM}$ & M$_{\rm DM}$ (\Msun) & dx (ckpc) & dx$_{\rm phys}$(kpc,z=5) & $x_{\rm HI}(z=5)$  &Comment \\
    \hline
    L04N0256 & 1 & 1 &  $256^3$ & $4.07 \times 10^5$ & 23.08 & 3.85 &$5.59\times 10^{-6}$  & Fiducial \\
    L04N0512 & 1 & 1 & $512^3$ & $6.21 \times 10^4$ & 11.04 & 1.84 &$3.94\times 10^{-6}$  &dx$_{\rm fiducial}$/2 \\
    L04N1024 & 1 & 1 & $1024^3$ & $7.77 \times 10^3$ & 5.52 & 0.92 &$2.45\times 10^{-6}$  & dx$_{\rm fiducial}$/4 \\
    L04N1024c005 & 0.05 & 0.05 & $1024^3$ & $7.77 \times 10^3$ & 5.52 & 0.92 &$3.23\times 10^{-5}$  & dx$_{\rm fiducial}$/4 \\
    \hline
  \end{tabular}
  \caption{Summary table of the simulations of our resolution study, giving the number of particles, the dark matter particle mass, the spatial resolution in comoving kpc, and the resulting neutral fraction obtained at z=5. All simulations  are performed with the full speed of light, except for L04N1024c005 which is a RSL simulation with \cpr = \cch = 0.05.}
  \end{center}
\end{table*}

The boosts in neutral fraction with resolution we find are similar to those obtained by \cite{rahmati2018} in their resolution study.
Our most resolved simulation is similar to the L050N0512 and L025N0256 cases of \cite{rahmati2018} in terms of spatial resolution, and reaches the sub-kiloparsec physical spatial resolution constraint stated by these authors even at z=5. With this high resolution, which is in principle high enough to resolve LLSs, our neutral fraction is $\sim 0.35$ dex higher, i.e. a factor of about 2 larger than our fiducial runs. This is a rather small impact, taking into account the enormous increase in computational resources required to perform this high resolution run. We cannot rule out that further increases in resolution would not push the neutral fraction to even higher values, but such simulations would in any case be tremendously expensive.

We also show the result of a high resolution RSL simulation with \cpr=\cch=$0.05 c$ in Fig. \ref{f:res}, along with its summary results in Tab. 1. Even with this high resolution set-up, reducing the speed of light still has a strong impact on the post-overlap neutral fraction.

\section{Conclusions}
\label{s:conclusions}
We have investigated the impact of the RSL approximation on the properties of the IGM using fully coupled RHD RAMSES-CUDATON simulations. We find that reducing the speed of light by a factor 5 (20, 100) leads to overestimating the post-reionization neutral hydrogen fraction by the same factor $\sim$5 (20, 100, respectively) with respect to our reference simulation using the full speed of light. We consider the simple analytical expression for a hydrogen gas in photoionization equilibrium to show that the error is driven by the hydrogen - photon chemistry. Reducing the speed of light has the same effect as artificially reducing the reaction cross section or reducing the photon density, and results in underestimating the ionizing intensity ${\rm J_{21}}$. We confirm this interpretation by running additional simulations using a reduced speed of light in the propagation module, but keeping this time the true speed of light in the chemistry module. With this set-up, dubbed DSL because of the simultaneous use of two different speeds of light, the post-reionization neutral hydrogen fractions converge to their full speed of light value, which validates our explanation.
In moderately ionized regions ($x_{\rm HI}>0.1$) and for a given photon density, the dependency between $x_{\rm HI}$ and the adopted speed of light $c$ becomes less strong, and in quasi-neutral regions, the equilibrium neutral fraction may become insensitive to the adopted speed of light.

It is important to note that the results presented in this work apply only to those implementations of RSL in which the background is not followed differently from the amplification around sources and does not affect those simulations using the implementation of \cite{gnedin2001}.
However, for the nonetheless very popular class of RT implementations on which  we focussed, we exposed an overlooked, fundamental issue of the RSL approximation, which should be kept in mind when interpreting results obtained in this framework.
We caution that these results are important in the post-overlap epoch and not necessarily beforehand.

We also performed a resolution study and show that spatial resolution may affect the post-overlap volume-weighted average neutral fraction. We increased spatial resolution beyond a cell size of 1kpc physical, so as to better resolve LLSs. At the highest spatial resolution we were able to achieve, we still observed a strong impact of the chosen speed of light on the post-overlap neutral fraction. Hence, spatial resolution within the limited regimes we were able to explore (up to four times our fiducial spatial resolution, i.e. an increase of 64 in mass resolution) does not significantly affect our conclusions.





As a major goal for the community, it is of prime importance that RHD codes achieve consistent predictions of the volume-weighted average neutral fraction, during and after the end of reionization, because this quantity is often used to calibrate such simulations, although we hope it will be increasingly replaced by a calibration using opacities when possible. Also, since the neutral hydrogen fraction is such an important parameter, we urge all colleagues and authors to always show this quantity rather than, or along with, the average ionized fractions, including after overlap.
Moreover, authors should be very careful to spell out properly the type of RSL approximation used, and its implementation in the photon propagation steps and in the thermochemistry module. This information is essential to understand and gauge the significance of the flurry of studies published these recent years and ongoing.





\section*{Acknowledgements}
The simulations used in this study were performed on CSCS/Piz Daint (Swiss National Supercomputing Centre), as part of the ``Shining a light through the dark ages'' PRACE allocation obtained via the 16$^{\rm th}$ call for PRACE Project Access. Auxiliary simulations were performed at the HPC center of the Strasbourg university. PO acknowledges funding from the Agence Nationale de la Recherche through the project ORAGE. The author thanks D.~Munro for freely distributing his Yorick programming language\footnote{http://www.maumae.net/yorick/doc/index.html}

\appendix

\section{Over-ionization in dual speed of light}
\label{s:overi}
\begin{figure}[b]
  \includegraphics[width=0.9\linewidth]{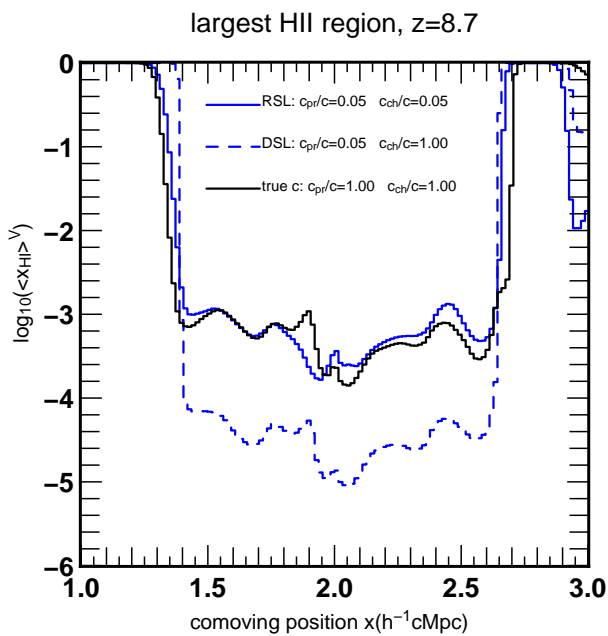}
  \caption{Neutral hydrogen fraction profile around the most massive halo of the simulation at z=8.7.}
  \label{f:xHIprofile}
\end{figure}

In this appendix we look at the ionization state pre-overlap in DSL. Indeed, a slower propagation speed artificially increases the photon density around sources as compared to the true speed of light case. In RSL, this is exactly compensated by a reduced speed of light in the photoionization rate, and this is why RSL yields correct neutral fractions for Stromgren spheres, as shown by \cite{rosdahl2013}. Conversely, in DSL, where the photoionization rate uses the true speed of light, the increased photon density produces over-ionized Stromgren spheres. This is illustrated in Fig. \ref{f:xHIprofile}.

\section{Star formation rates in RSL and DSL}
\label{s:sfrrsldsl}

\begin{figure*}[t]
  \begin{center}
  \includegraphics[width=0.45\linewidth]{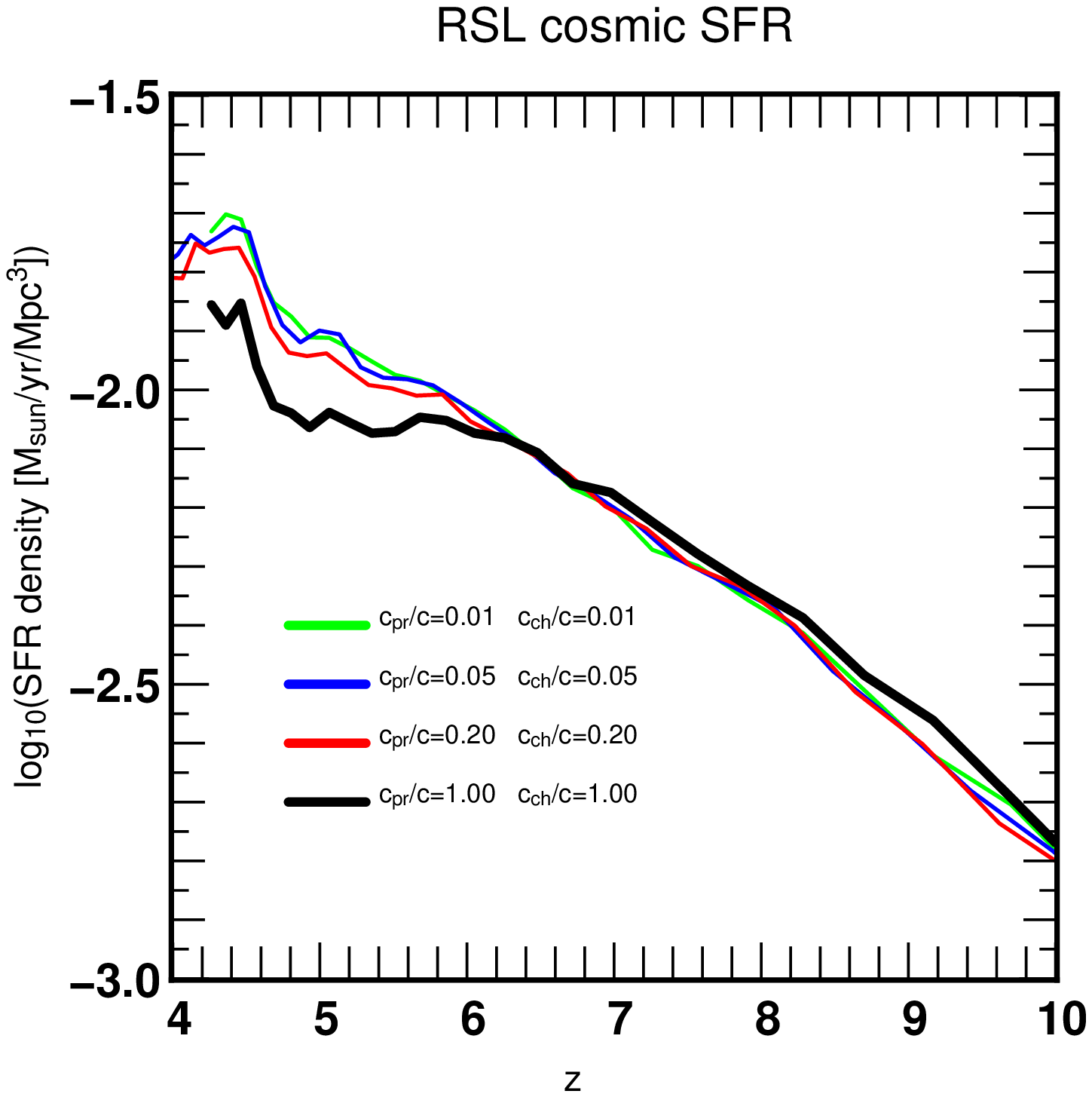}
  \includegraphics[width=0.45\linewidth]{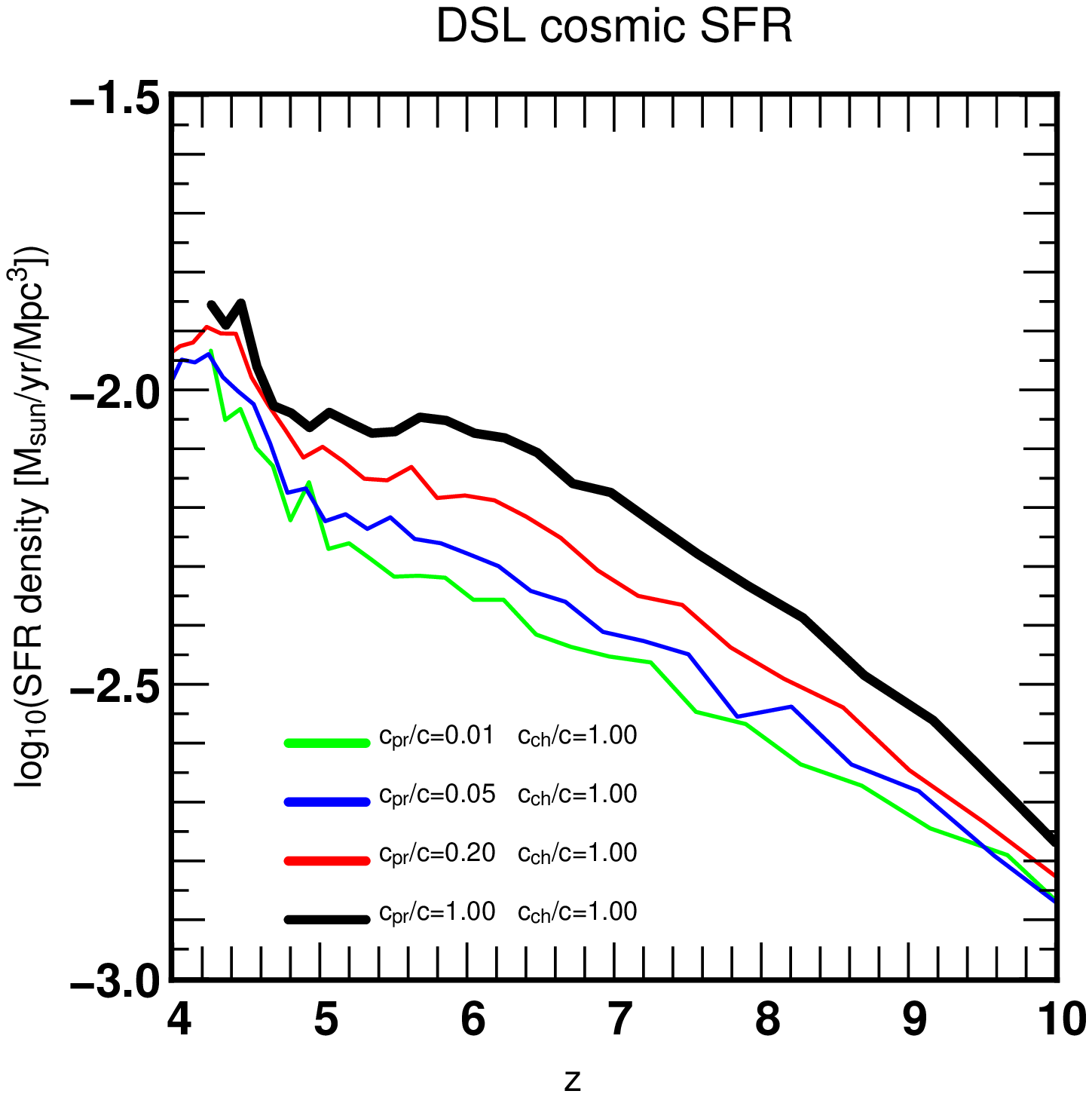}
  \end{center}
  \caption{Cosmic SFRs for the RSL (left) and DSL (right) runs.}
  \label{f:sfrs}
\end{figure*}

We showed in Fig. \ref{f:DSL} that the lag in reionization redshift is even worse in DSL than in RSL. All DSL runs reionize significantly later than their RSL counterpart. The most likely culprit for this is a deficit of star formation in DSL runs compared to the full speed of light and RSL runs. This is shown in Fig. \ref{f:sfrs}. In RSL, the SFR of the box is rather close to the full speed of light SFR all the way to the overlap and starts to diverge afterwards, where SFRs split into a sequence ordered by decreasing speeds of light: a slower speed of light results in a higher SFR after overlap.
{
  This result can be related to the findings of \cite{dawoodbhoy2018}, who showed using the CoDa I simulation \citep{codaI} that star formation in low-mass haloes was reduced by the reionization of their local patch. In our experiment, the increasingly late reionization histories obtained in RSL would therefore trigger this suppression at later times, resulting in the difference found between the full speed of light SFR and RSL SFRs.

Conversely, the DSL runs yield opposite results in both aspects: the SFRs differ { before} overlap and seem to converge { after} overlap, and the sequence of SFRs is ordered in decreasing speeds of light: a slower speed of light results in a lower SFR. This deficit in SFR, in turn, results in a slower reionization history, in particular with the smallest speed of light considered.

This requires a different explanation. We showed in Fig. \ref{f:xHIprofile} that DSL yields over-ionized Stromgren spheres, which may hamper cooling and subsequent star formation. Indeed, in pure hydrogen chemistry, neutral hydrogen thermal excitation, and subsequent line emission are the main cooling processes between $10^4-10^{5.5}$ K \citep{katz1996}; the cooling rate involves the product of the neutral hydrogen and electron densities $\rho_{\rm HI} \times \rho_{\rm e}$, which is maximum at ${\rm x_{HI}}=0.5$ in a hydrogen-only plasma, and decreases for lower neutral fractions.
}



We note, however, that the rather strong impact of $c$ on our SFRs, is exaggerated by the relatively low density threshold for star formation we used. Indeed, Fig. \ref{f:cobraRSL} shows that our density threshold for star formation is in the range where the neutral fraction is strongly affected by $c$. If we could use as star formation threshold a comoving overdensity of $10^4$, for instance, where gas is more likely to be neutral according to Fig. \ref{f:cobraRSL}, it is possible that the impact of $c$ on the SFR would be less important. However, with such a threshold star formation would be extremely rare in the set-up used in this work, and this would prevent us from obtaining a reasonable reionization history. This approach is therefore not viable here, but may be in a very high resolution simulation.


\bibliographystyle{mn2e}
\bibliography{mybib}
%

\label{lastpage}
\end{document}